# PERSA+: A Deep Learning Front-End for Context-Agnostic Audio Classification

Lazaros Vrysis, Iordanis Thoidis, Charalampos Dimoulas, and George Papanikolaou.

*Abstract*—Deep learning has been applied to diverse audio semantics tasks, enabling the construction of models that learn hierarchical levels of features from high-dimensional raw data, delivering state-of-the-art performance. But do these algorithms perform similarly in real-world conditions, or just at the benchmark, where their high learning capability assures the complete memorization of the employed datasets? This work presents a deep learning front-end, aiming at discarding detrimental information before entering the modeling stage, bringing the learning process closer to the point, anticipating the development of robust and context-agnostic classification algorithms.

## I. INTRODUCTION

MANY contemporary audio applications rely on semantic audio analysis workflows that even run on mobile computing terminals [1]. Therefore, the improvement of audio classification algorithms has become a necessity. In recent years, *Convolutional Neural Networks (CNNs)* have set the standard on audio semantics, surpassing conventional methods which employed hand-crafted feature extraction as the front-end and a classifier as the back-end [2]. In the meantime, the process of temporal feature integration introduced additional engineering steps, attempting to capture the temporal dependency between successive feature observations, delivering performance improvements [3]. Now, *CNNs* have been applied to diverse *machine learning (ML)* tasks, delivering state-of-the-art performance, enabling the construction of models that learn hierarchical levels of features from high-dimensional data, i.e. spectrograms or even raw waveforms [4]. One question that arises, though, concerns the generalization of those algorithms, i.e. their ability to maintain robustness when encountering acoustic samples outside the deployed datasets [5].

The motivation behind this work emerged after the experimentation with a deep learning model on a speaker recognition problem and the empirical observation of a rather strange behavior. At some point, while developing the model, recognition accuracy reached a surprisingly high value (above *95%* for a ten-speaker task), while, normally, we were strangling to achieve a score of over *60%*. Unfortunately, after investigating the situation, it turned out that this unforeseen performance leap

L. Vrysis, I. Thoidis, C. Dimoulas, and G. Papanikolaou are with the Aristotle University of Thessaloniki.

occurred due to some abnormal data values: a bug in the audio processing library was producing artifacts on the produced spectrograms, and some *time-frequency (tf)* bins had the exact same -unique- values, based on the audio file they came from. These values were present in the high-frequency range of the spectrogram, forming a kind of a watermark. Therefore, any of the spectrogram slices that were used as input samples, contained a unique identifier. They were no more than *1%* of the total values that each sample consists of. Nevertheless, the neural network was capable of focusing on these values to complete the assigned task with ease, ignoring all the rest, useful, data. Obviously, this did not solve the classification problem with success, because the data was unique for each file and not for each speaker. A train/test subset formation flaw via sample shuffling did the trick: the model just learned all the watermarks for each speaker, which were present in both train and test subsets. It became apparent that a deep learning model may be easily trained in an unwanted way, just by memorizing insignificant amounts of meaningless data. It should also be mentioned that these *tf* areas had extremely low power, below *-90 dBFS*, way under the *dynamic range (DR)* of an audible, recorded sound. Therefore, according to our judgment, careful data curation and targeted preprocessing should always accompany deep learning algorithms to force them to focus on the target patterns and not just learn some random values.

This work presents a preprocessing method aiming at discarding detrimental information before entering the modeler, anticipating the development of more robust and context-agnostic classification algorithms.

## II. RELATED WORK

Hand-crafted feature extraction served the field of semantic audio analysis for many years [6]. Undoubtedly, deep learning brought higher accuracies, while the increased computational requirements are no more a barrier, as even the mobile computing devices offer enough computing power [7]. There is a variety of deep learning implementations for *General Audio Detection and Classification, Environmental Sound Recognition, Keyword Spotting, Speech Emotion Recognition*, and more. There is also a plethora of network architectures, including convolutional models [8], recurrent approaches [9], or even ensemble learning tactics [10]. That is, the emergence of more capable and efficient deep learning models for audio semantic analysis should not overlook the need for audio-specific pre- and post-processing strategies, in applications where interpretability and generalization are crucial. In this direction, some considerations are expressed in the following paragraphs.

First, end-to-end learning has been adopted from the image analysis field and most of the early studies modified *2D* topologies for image recognition, adapting them to audio classification tasks, without taking into consideration the particularities of semantic audio analysis [11]. Audio waveforms are usually converted to *tf* representations, such as spectrograms. Recently, many approaches are characterized as end-to-end and process *1D* data, employing filters with size and stride that are comparable to the window and the hop size that are used by the *Short-Time Fourier Transform (STFT)* [12]. Newer designs make use of smaller filters (e.g. *10* sample-long filters at *16kHz*) [13]. The sample-level *CNN* is an excellent example of this approach, specifying filters with very small granularity in time for all convolutional layers, achieving decent performance [14].

Second, network architecture is always in the spotlight, as it may have an impact on the effectiveness of a deep learning model. *U-Net* [15] and *Inception* [16] are two examples of important milestones in the development of *CNN* classifiers, featuring a more complex architecture than the early approaches. *SincNet* is a network dedicated to audio semantic recognition that is based on parametrized sinc functions, which implement band-pass filters that only learn low and high cutoff frequencies [17]. Nevertheless, no noticeable gains have been observed when deploying these models [18], a consideration that is in accordance with the results of similar works that employed significantly larger datasets and flagship *CNN* architectures [19].

Third, there is evidence that *CNNs* are prone to overfitting and gaining a dataset bias [20]. Deep architectures require larger amounts of data for a meaningful training, in contrast to conventional learning models. For this reason, various data augmentation techniques are typically used. Furthermore, *CNNs* are subjected to data deformations, such as loudness variations, additive noise, etc. [21]. In past studies, the objective was to design robust and context-resistant techniques that could be used in real-world applications. Nowadays, the trend is to train the deep learning models with huge amounts of data to make the most of them. Approaches like *MVA* [22] and *RASTA* [23] that, for example, give noise robustness in *Automatic Speech Recognition (ASR)*, are profound for sure. This is crucial since, in many applications, a trained *ASR* system needs to carry out recognition in a variety of everyday acoustic environments. It seems that this is not the case in the deep learning era. Many studies do not specify any pre-processing method to deal with inputs of different scaling or contexts, although it is obvious that data curation has a major impact on the performance of deep learning algorithms [24]-[25]. Ambiguous normalization strategies are deployed in many works [11], while there are cases where data normalization is based on the whole dataset, probably getting a hint of the test data [14].

For these reasons, we believe that state-of-the-art deep neural networks should be accompanied by well-defined pre-processing protocols so that the model adapts to the particularities and constraints of the audio domain. For instance, *Pons et al.* [26]-[27] have conducted exceptional research on deep network designs for audio analysis. They follow a two-part model decomposition with the front-end being the part of the model that interacts with the input signal to map it into a latent space, and the back-end predicting the output, given the representation obtained by the former. This approach is highly appreciated, as it can be considered more methodical, better facilitating the optimization of data pre-processing flows. Some works do indeed emphasize on data normalization, aiming at improving the robustness of the systems, deploying sophisticated front-ends along with context-adaptive recognition strategies [28]. A notable front-end of this kind is *PCEN*, which improves robustness to loudness variations [29]. This can be applied in far-field recording conditions, where signals are attenuated due to the distance. The key feature of *PCEN* is the use of dynamic compression that is based on automatic gain control, replacing the widely used static, log compression. On large evaluation sets that include noisy data and far-field recording cases, *PCEN* improves recognition performance [29]. Additionally, it can be implemented in real-time and be distributed across sensors, preserving the locality structure of harmonic patterns along the mel-frequency axis [30]. Nevertheless, there are occasions that *PCEN* does not manage to deliver the expected results, and a combination of methods is required for optimal operation [28].

These findings were taken into consideration for proposing a preprocessing protocol targeted to deep learning architectures, aiming at homogenizing data for eliminating the impact of the acoustic context on audio recognition and increasing the robustness of the models. That is, instead of augmenting data for simulating different versions of a particular acoustic event, we try to shape each single pattern to be as representative as possible.

III. PER SAMPLE ENERGY NORMALIZATION

As highlighted, it is probable for a classification algorithm to perform worse outside in-vitro conditions [31]. A factor that may impact real-world performance can be the variations of the signal power [2]. We consistently try to avoid any input-gain bias to our algorithms -except when it is required- by excluding energy-based audio features in standard machine learning approaches [32]. This principle is followed on the *CNNs* as well, adopting the *PER-SAmple (PERSA)* normalization strategy [2]. The latter ensures zero bias to input gain, eliminating this level of freedom. In brief, *PERSA* specifies the subtraction of the log-mean value on any *2*-dimensional *tf*-sample, before entering the convolutional network. The length of the sample should be relatively large (i.e. from *500 ms* to *5000 ms*) for avoiding changes in the dynamics of the signal. Hopefully, sample lengths of the same magnitude are preferred in most deep learning audio recognition tasks.

IV. CONTEXT-AGNOSTIC FRONT-END PROCESSING

The *PERSA+* front-end comes as an extension of *PERSA*, expecting to favor performance on noisy recordings and bring consistency on diverse acoustic contexts. It is based on the hypothesis that: *(a)* parts of audio data that lay into the lower *DR* may not be crucial for audio classification and *(b)* it could be better for these parts not to be routed to the modeler because, instead of useful data, it is more probable to contain noise dis-tributions that can be watermarked. The developed method was built on these assumptions, specifying noise injection to the data before entering the classification unit (both in train and test phases), expecting that it will handle noisy data more accurately but without compromising performance in high *Signal-to-Noise Ratios (SNRs)*. *PERSA+* specifies three core processing steps: *(a) Noise Injection, (b) Log Compression,* and *(c) Mean Subtraction*. Let us describe these in more detail.

Let the *k-th* sample of the signal have the form of a *2D* two-dimensional *tf* representation comprising of *L* time frames and *M* frequency bands, so as a *tf* bin can be noted as *s[i, j]* with $\{i \in Z \mid 0 \leq i < L\}$ and $\{j \in Z \mid j \in 0 \leq j < M)\}$.

### A. Noise Injection

First, a *tf* representation of a randomly generated noise segment *(N)* is generated (i.e. pink noise). Let $p_s$ and $p_n$ denote the power of the signal and noise segments, respectively. The level of noise injected to $S_n$ is determined by the parameter $q$ (in *dB*).

$$S_n[i,j] = \left(S^2[i,j] + N^2[i,j] \frac{p_s}{p_n \cdot 10^{\frac{q}{10}}}\right)^{1/2} \quad (1)$$

### B. Log Compression

Afterward, the log function with base *10* is applied to $S_n$:

$$S_{ln}[i,j] = log_{10} S_n[i,j] \quad (2)$$

### C. Mean Subtraction

Then, the mean value of the *tf* slice is subtracted:

$$S_{persa+}[i,j] = S_{ln}[i,j] - \frac{1}{L \cdot M} \sum_i^L \sum_j^M S_{ln}[i,j] \quad (3)$$

At this point, it should be noted that the final processing step could alternatively employ the division of each *tf* bin by the mean energy of the whole sample, before applying the logarithmic function. The two approaches (log - mean subtraction vs. mean division - log) have the same impact on the shape of the final spectrotemporal sample with a slightly different offset. These two variations were tested and the former yields slightly better -and more consistent- results, directing the final decision. Finally, a small amount of random gain *(±3 dB)* is applied for differentiating similar *tf* samples, making the data more meaningful for training.

It is obvious that *PERSA+* depends on the parameter *q*. High values (e.g. *60 dB*) virtually make no difference when compared to the standard method *(PERSA)*, whereas lower values (e.g. *12 dB*) significantly affect the input data, as a relatively high noise floor is set. This may bound the performance of the classification system, but with the promise for better performance under low *SNR* scenarios. All these concerns were taken into consideration for designing the appropriate experimental setup.

## V. EXPERIMENTAL SETUP

### A. Overview

The experimental setup employs the comparison of various audio preprocessing techniques, under typical classification tasks, while a special dataset has been developed to simulate low *SNR* conditions as well. A common *2D CNN*, applied on mel-spectrogram audio representations, was selected as the core classification pipeline. Technically, *Python* was used for delpoying the experimental setup with *librosa* [33] and *Keras* facilitating the feature extraction and deep learning procedures [34], respectively.

### B. Datasets

The performance of the competitive methods was mainly evaluated on a *3-class* classification task, according to the *Speech/Music/Other (SMO)* taxonomy. The *LVLib-v1* dataset was deployed as a baseline. To conduct additional experiments, the *LVLib-v3* was also employed. This dataset was generated after applying random gain *({0, -10, -20, -30} dBFS)* on *LVLib-v1*, for each fold and class, respecting a *3-fold* setup. Following a similar philosophy, the *LVLib-v4* brings further signal degradations. In specific, audio waveforms are contaminated with additive interference signals. The interference follows the *SMO* taxonomy and was added according to a specific protocol: 1[st] fold - *Music* is contaminated with *Speech*, *Speech* with *Music* and *Other* with machinery noise, 2[nd] fold - *Music* is contaminated with *Other*, *Speech* with machinery noise and *Other* with *Music*, 3[rd] fold - *Music* is contaminated with machinery noise, *Speech* with *Other* and *Other* with *Music*. The *SNR* of the additive noise follows a *Gaussian* distribution *(9 ±3 dB)*. When *LVLib-v4* is used, the algorithms are trained on *LVLib-v3* with the test fold substituted by the corresponding fold of *LVLib-v4*. Moreover, a combination of *LVLib-v1* and *v4* sets was used, simulating mixed conditions. All *LVLib* datasets are publicly available at m3c.web.auth.gr/research/datasets/. A *3-fold* data splitting is recommended, facilitating the direct comparisons between the results of different studies. One more task was conducted on a *10-class ESR taxonomy*, using the well-known *UrbanSound8k* dataset [35].

### C. Competing Front-ends

Five different front-ends were tested: log-compression *(LOG)*, log-compression with amplitude augmentation *(LOG-AU)*, log-compression with thresholding *(LOG-T)*, *PERSA*, *PERSA+*, and *PCEN*. The *LOG* method employs the log transformation of the spectrogram power. *LOG-AU* additionally brings magnitude augmentations for overcoming any level bias on the classifier. That is, random gain is applied before the data entering the back-end, according to a uniform distribution with the *a* and *b* parameters set to *-30* and *30 dB*. Next, *LOG-T* specifies a per-sample maximum normalization with logarithmic compression, compressing the *DR* of the signal, as denoted by the following equation:

$$S_{LOG-T} = log_{10}\left(S + 10^{\frac{c}{20}}\right) \quad (4)$$

where parameter *c* corresponds to the desirable DR (in dB). Finally, The *PCEN* front-end was deployed using the default *librosa* setup [33].

### D. Backend

Based on former results, the simple *VGG-like* architectures show adequate performance in all circumstances [18]. Therefore, a decision to use a lightweight backend of this type with roughly *100k* parameters was made. It consists of four successive *CPD* blocks (comprised of successive *Convolutional, Pooling,* and *Dropout* layers), a *Global Average Pooling*, two *Fully Connected*, and an intermediate *Dropout* layer. Filter-size was set to *3x3* globally, while the number of filters was set to *16*, *32*, *64,* and *128* for the four convolutional layers. The *pooling* size was set to *2x2* for all layers of this kind. The *fully connected* layers had *32* and *[number of classes]* neurons respectively. The rest of the parameters are the *ReLU* as the activation function for all intermediate layers, *SoftMax* for the output layer, *Categorical Cross-Entropy* as the loss function, and *Adam* as the optimizer. The *dropout* rate was set to *20%*.

## VI. RESULTS

First, for the *PERSA+* and *LOG-T* algorithms a hyperparameter optimization procedure was carried out. Figure 1 demonstrates the performance of the two methods on *LVLib-v3*, with respect to the *q* and *c* parameters. The values of *9 dB* and *30 dB* were selected respectively, aiming at maximizing the effect of the methods, without causing significant performance degradation on clean data. Table I presents the accuracy of *PERSA+* when deployed on *LVLib-v4*, under variable *SNR* versus different *q* values. The relation between the *SNR* and *q* parameters in terms of classification performance underpin the impact of the noise injection strategy when dealing with noisy data. Table II shows the scores for the competitive methods on three different tasks. All methods perform well on the *LVLib-v1* dataset *(85.9 ±0.9%)*, with these that theoretically are more robust *(LOG-T, PCEN, PERSA+)* having slightly lower ratings *(85.2 ±0.6%)* than the "standard" ones *(LOG, LOG-AU, PERSA - 86.6 ±0.7%)*. On the *LVLib-v3*, which employs input-gain variations, the *LOG and LOG-AU* methods fail with a performance reduction of about *6.5%* and *4.5%,* respectively, whereas all other approaches keep their ratings (mean performance drop of *0.2%*). On the *LVLib-v4*, which has noisy data, the *PERSA+* shines. A wide performance gap is evident between the first the rest of the methods, with the scores being *75.8%*, *73.3%*, *73.9%* for the *PERSA+*, *LOG-T,* and *PCEN* front-ends, respectively.

Table III depicts performance ratings for all methods on two additional tasks. The proposed method prevails on the *LVLib-v1* and *v4* combination, while shows slightly lower accuracy on *UrbanSound8k* than the best performing method *(PERSA)* of about *1%*. In short, *PERSA+* does indeed manage to handle low *SNR* scenarios better than all competitive methods, without compromising performance on normal conditions.

## VII. CONCLUSION

In general, experimental results prove that *PERSA+* verifies its design assumptions, matching the performance of *PCEN*. The proposed front-end layer favors robustness and universality. At this point, it should be stated what exactly *PERSA+* is meant to do: (a) it can simulate noisy conditions that are closer to real-world scenarios, (b) it can lead to more representative results for the classification algorithms and unveils their true potential to perform equally in different contexts, (c) it can be used for avoiding misleading performance ratings that may be the outcome of overfitting, (d) it can be used for rapid experimentation on small datasets as it performs consistently across different classification tasks, (e) it delivers higher classification accuracy on noisy data, with minimum losses on clean data, and (f) it is probable to perform slightly worse on most occasions, especially on the test bench.

Summing up, a neat front-end for audio classification was presented, while a solid evaluation was executed. Positive results were observed and the method seems potent. Nevertheless, further tests have to be made on more benchmark datasets. The keyword-spotting task is mainly under consideration for a more complete evaluation, while further modifications and improvements, like the deployment of alternative *tf* representations, time-domain operation, and signal and noise weighting are under investigation.

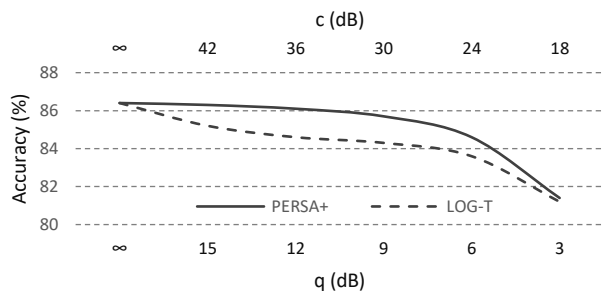

Fig 1. Classification performance of PERSA+ and LOG-T front-ends on LVLib v3 with respect to q and c parameters.

TABLE I
CLASSIFICATION ACCURACY (%) FOR THE PERSA+ FRONT-END WITH RESPECT TO PARAMETER Q ND NOISE CONDITIONS

|   |   | SNR | | | | | |
|---|---|---|---|---|---|---|---|
|   |   | ∞ | 15 | 12 | 9 | 6 | 3 |
| q | ∞ | **86.4 ±3.9** | 79.5 ±7.3 | 74.6 ±5.6 | 70.3 ±5.4 | 65.1 ±5.7 | 59.2 ±7.5 |
|   | 15 | 86.3 ±4.6 | 81.2 ±6.3 | 78.5 ±5.6 | 72.0 ±6.6 | 67.1 ±3.5 | 61.2 ±5.3 |
|   | 12 | 86.1 ±7.3 | **82.4 ±6.3** | **79.5 ±4.4** | 75.6 ±5.4 | 68.4 ±5.4 | 59.3 ±4.1 |
|   | 9 | 85.7 ±4.9 | **83.6 ±5.4** | **80.3 ±6.9** | 75.8 ±5.4 | **69.0 ±4.9** | 61.3 ±6.1 |
|   | 6 | 84.6 ±5.8 | 81.7 ±5.9 | 78.3 ±3.2 | **75.1 ±5.3** | **68.5 ±6.5** | **63.1 ±3.1** |
|   | 3 | 81.4 ±3.5 | 80.8 ±3.5 | 78.3 ±3.3 | 74.2 ±5.2 | 66.2 ±6.3 | **62.5 ±7.0** |

Values above *-1%* of the higher value for each column are highlighted.

TABLE II
FRONT-END PERFORMANCE ACROSS THE LVLIB SMO DATASETS

|   | LVLib-v1 *baseline* | LVLib-v3 *random gain* | LVLib-v4 *random noise* |
|---|---|---|---|
| LOG | 86.2 ±6.0 | 79.8 ±0.5 | 58.9 ±9.9 |
| LOG-AU | 85.9 ±5.5 | 81.4 ±0.4 | 61.0 ±8.6 |
| LOG-T | 85.3 ±5.0 | 85.2 ±5.1 | 72.3 ±2.1 |
| PERSA | **87.6 ±4.1** | **86.4 ±3.9** | 73.3 ±5.4 |
| PERSA+ | 85.9 ±5.8 | **85.7 ±4.9** | **75.8 ±5.4** |
| PCEN | 84.4 ±6.4 | 85.1 ±6.5 | 73.9 ±2.4 |

Values above *-1%* of the highest value in each column are highlighted.

TABLE III
FRONT-END PERFORMANCE ON ADDITIONAL TASKS

|   | LVLib v1+v4 | UrbanSound8k |
|---|---|---|
| LOG | 69.9 ±0.7 | 71.1 ±6.6 |
| LOG-AU | 71.1 ±1.6 | **71.9 ±5.2** |
| LOG-T | 78.0 ±2.8 | 69.7 ±5.1 |
| PERSA | 79.8 ±1.1 | **72.2 ±5.9** |
| PERSA+ | **81.8 ±3.1** | 71.4 ±4.6 |
| PCEN | 79.5 ±3.1 | **71.8 ±2.5** |

Values above *-1%* of the highest value in each column are highlighted.